# INTEGRATED, RELIABLE AND CLOUD-BASED PERSONAL HEALTH RECORD: A SCOPING REVIEW

Jesús Romero, Pablo López, José Luis Vázquez Noguera, Cristian Cappo, Diego P. Pinto-Roa and Cynthia Villalba

Facultad Politécnica, Universidad Nacional de Asunción, Asunción, Paraguay

## ABSTRACT

*Personal Health Records (PHR) emerge as an alternative to integrate patient's health information to give a global view of patients' status. However, integration is not a trivial feature when dealing with a variety electronic health systems from healthcare centers. Access to PHR sensitive information must comply with privacy policies defined by the patient. Architecture PHR design should be in accordance to these, and take advantage of nowadays technology. Cloud computing is a current technology that provides scalability, ubiquity, and elasticity features. This paper presents a scoping review related to PHR systems that achieve three characteristics: integrated, reliable and cloud-based. We found 101 articles that addressed thosecharacteristics. We identified four main research topics: proposal/developed systems, PHR recommendations for development, system integration and standards, and security and privacy. Integration is tackled with HL7 CDA standard. Information reliability is based in ABE security-privacy mechanism. Cloud-based technology access is achieved via SOA.*

## 1. INTRODUCTION

Nowadays healthcare delivery model is changing from institution-centered to a more patient-centered model[1]. The past fragmented patient health information [2] is changing to patient-centered that enables patients to store and access to their health information ubiquitously in a personal health record (PHR) [1].PHR is defined as a set of computer-based tools that allow people to access and coordinate their lifelong health information and make appropriate parts of it available to those who need it[3,4]. There is not agreement or standard on what information a PHR should store[5]. Some shared information supported by PHRs are problem lists, procedures, major illnesses, provider lists, allergy data, home-monitored data, family history, social history and lifestyle, immunizations, medications and laboratory tests[6].

When patients assist to different healthcare centers, their health information remains distributed across the healthcare centers visited. An integrated health record will provide a global and complete view of a patient health state that can lead to a better decision support by physicians during consultation, or even in emergency situations[7].

Cloud computing is a model for enabling ubiquitous and on-demand access to a shared pool of services [8]. Ubiquitous access to information not only benefits to patients but also, to healthcare professionals. Due to the high level of mobility physicians experience, ubiquitous access to relevant and timely patient data would help to make critical care decisions [9], even save lives [7].Ubiquity ease access to patient records, it could lead to privacy issues and unauthorized access.

PHR stores sensitive patient health information, therefore, access to PHR data must comply to privacy policies defined by the patient. In this work, we use the term reliable to express the accomplishment of this privacy policies. Privacy is recognized as the most sensitive aspect of e-health record systems [10] and must be achieved through an appropriate mechanism.

Security and privacy of health data are one of the major concerns in e-health. Due that Cloud Server Provider generally is a third party component, health data should be securely stored to protect the privacy. Cloud servers are considered to be semi-trusted, because they will not actively try to get the information stored, but they make for example traffic analysis, which may expose data. Solutions in the literature include encryption of data before outsourcing to cloud, access control, and party identifications to valid who is getting access.

Our scoping review question is: What are the current implemented or proposed PHR systems that achieve next characteristics: integrated, reliable and cloud-based?

The term *integrated* is referred to unified information that can be shared to authorized healthcare stakeholders (e.g., physicians, nurses, medical organizations).

PHR systems store sensitive patient information that could help save lives when used in emergency situations. Because of this, patient's information should always be consistent to be *reliable* health stakeholders.

In this study we consider ubiquity a feature that should be take into account for PHR system design. In order to ease information access to health information from portable devices (as smartphones and tablets), cloud computing is a technology that provides ubiquity by definition [8]. Many articles do not consider ubiquity as a system design goals. However, same articles address systems that are *cloud-based*. Thus, we decided to search term *cloud* to fulfil ubiquity.
Our main focus in this review is to identify PHR implemented or proposed PHR systems that apply three characteristics: integrated, reliable and cloud-based.This article is organized as follows. In Section 2 we explain the methodology applied in this scoping review. Then, in Section 3 results are detailed. In Section 4 articles statics figures and discussions are pointed out. Finally, in Section 5 conclusion is explained.

## 2. METHODS

As a scoping review, our process consisted in three stages as follows: searching phase, iterative filtering, and analysis.

In the first stage *searching phase*, we defined a search string and chose the database. As our review question stays, we focused on searching articles related to PHR that cover three characteristics (integrated, reliable and cloud-based). For this purpose, we defined the search string as "PHR AND (integration OR standard) AND privacy AND cloud". *PHR*is a keyword for searching*personal health record system.Integration or standard* refers to term "integrated". Information integration is often base in the used of standards. To have reliable information, *privacy* must be defined into healthcare system. Last term "cloud" constitute the search string for cloud computing. The database resource used was *Google Scholar*.

In the second stage of*iterative filtering*, we contemplated three filtering sub-stages as follows:

- Database string search: 1410 articles were found as a result of the search string. This was obtained by unchecking patents and citations filter options in *Google Scholar* page.
- Filtering by global screening: 189 papers were filtered based on a global screening. By just reading title, abstract and keywords, we judged if papers contemplated the main research focus: PHR systems that cover three characteristics (integrated, reliable and cloud-based). Many articles found centered in one, or sometimes, two of the characteristics. Each of the characteristics searched can encompass a deep research study in PHR area, and just one topic could be addressed in an article. Thus, articles that are related to the review question partially were also included.
- Filtering by complete reading: we included 101 articles in our final synthesis. Duplicated and redundant articles were excluded. Moreover, papers published in languages other than English, uncompleted works were considered as non-scientific and excluded as well. Articles that were not related to PHR with at most one of the characteristic (e.g. cloud computing topics or internet security) were not in the scope of this review.

The whole scoping review process is summarized in Figure 1. The two stages explained in this section are detailed in the first top bands. The bottom band addresses the third stage of articles analysis. This is detailed in next section.

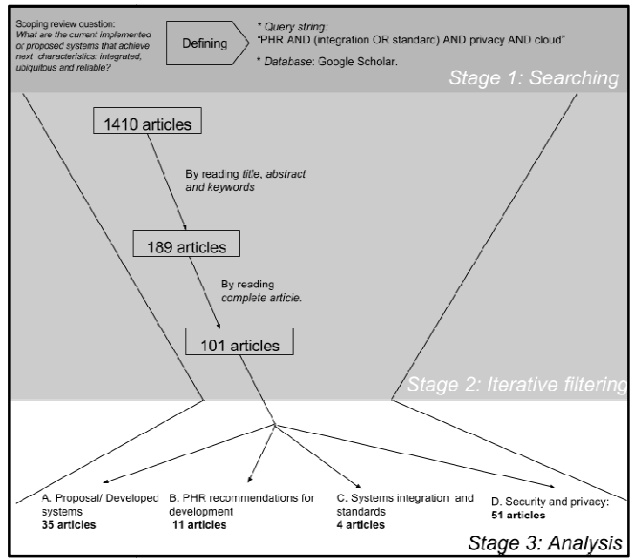

Figure 1. Scoping review process summary

## 3. RESULTS

After the analysis of 101 articles, we identified four research main topics. Each article was read again and made a synthesis in a text document where each characteristic (integrated, reliable and cloud-based) was sketched. Later, by reading sketched summaries of each work, articles with similar topics were grouped. Final research topics are cited below:

A. *Proposal/Developed systems:* articles that focus on PHR systems were included. Authors address system architecture technology, and, in some of them, tests were carried out to their proposal.

B. *PHR recommendations for development:* this topic gathers articles that analyses and identifies broad issues to certain themes in PHR system design. Based on theses common issues, authors provide tips and recommendations to PHR system design and development.
C. *Systems integration and standards:* this includes articles that focus on integration of PHR. Architectures core integration and standards applied are detailed.
D. *Security and privacy:* this contains articles that explain the security and privacy mechanisms applied to PHR.

Some articles found address various topics and, hence, are referenced into various research topics.

## 3.1. Proposal/Developed Systems

This topic groups PHR articles that detail architectures, stack technology and healthcare scenarios where systems were implemented. Even though some articles included mention Electronic Health Record (EHR) as their main topic, we unified EHR with PHR system. EHR is an electronic version of the patient medical record kept by physicians and hospitals[11]. EHR integrates patient health information from different healthcare centers, and it is shared between them [12]. While EHR is controlled and intended for use by medical providers, PHR is controlled by patient [11].
Most shared approach in PHR architectures found is Service-Oriented-Architecture (SOA) [13,14,6,15,16,17]. This approach provides efficient, scalable, portable, interoperable and integrated IT infrastructure that is cost effective and maintainable [13]. Articles applied SOA focusing in a cloud architecture in a stack. The architecture stack is often compounded from three services: SaaS (Software as a Service), IaaS (Infrastructure as a Service), and PaaS (Platform as a Service). In SaaS layer services are implemented and accessed via a browser by users. In PaaS layer, web-services API are provided to clients that can take advantage of information stored. In IaaS layer, architectures components as databases, interconnectivity, and virtual machines are deployed. This last layer is the base for PaaS and SaaS. Cloud computing provides dynamically scalable computational resources, at low cost in a ubiquitous manner [7,17].

Doctors need, in critical moments, patients health information in real time, this is because right information at the right time saves lives [7]. Authors in [7]proposes a solution based on cloud computing implemented for hospital systems having as a result, a better management, high speed for the medical process, and increased quality of the medical services [7]. This model was applied into two key departments of a hospital: Pediatrics and Obstetrics, and Gynecology.

A system called MyPHRMachines applies a different architecture approach based on Virtual Machines (VM) and exposes its service over IaaS[18,19]. Patients access their PHR profile with a special*Remote Desktop Protocol*(RDP) client or a web-browser with JAVA. Patients can share part of their personal health information to stakeholders with a link. Storage is accessed only with a*Virtual Machines* (VM) images that are created on the server.

Other works considered that accessing to information in emergency time is crucial. In [17], authors developed a prototype system to integrate information between ambulances and healthcare centers. This work applied a BPEL (Business Process Execution Language) to retrieve, transform and stored patient information in virtualized cloud services.

Besides deploying their proposed systems in their own physical servers, numerous works relies on third services cloud. For instance, we cite some articles that uses Google Cloud Services [9,20,21,22,23], Amazon services[17], and Windows Azure [7].

By definition, PHR is patient-based information. This may be via external devices, or even patients uploading information by themselves. Some platforms focus on telemedicine, and provides modules that interact with sensors that achieve health information. For instance, sensor watches, mobile applications, online forms, among others [14,24].

A novel technology that has started to spread in database market is NoSQL storages. In [25] authors applies a NoSQL Hadoop in their architecture to enhance computation power and system scalability.

As a possible barrier to PHR system adoption may be the lack of internet rural zones. To alleviate this, authors in[4] proposes to store PHR information in a USB smart card or smartphone. Patients carry out their information with themselves and provide it to physicians during consultation.
Besides applying regular web browser or mobile access, other projects centered in applying ATM (Automated Teller Machine) kiosks to access their system [26,27]. A system named *HealthATM*integrates Google healthcare and Microsoft HealthVault Services [28] in a cloud-based approach.

In mobile area, we identified a group of articles that focus on mobile architecture platforms exclusively. In [9,22,23], a Greek cloud-based PHR system named *Nefeli* is explained. Using Google Cloud Messaging Services, the main goal in *Nefeli* design is to optimize the use of battery and mobile bandwidth for Android native mobile application accessing to the service [9].
In [29], a general model to manage healthcare hospital in Brazil named *uHospital* is presented. The system applies two characteristics: ubiquitous health (e.g. monitoring patients health anywhere and anytime) and ubiquitous healthcare (e.g. convenient services to patients that allow the clinical diagnosis). The model employs these two concepts to develop a PHR.

*uHospital*model primary goal is to analyse patient health information stored in their PHR to detect possible risks to their health. PHR information is stored in an ontology allowing the use of inferences to detect possible situations and defining involved risks.

In [30], a PHR system solution is proposed. The proposed PHR allows the exchange of patient data at the point-of-care using patient's mobile device.They propose a novel message protocol denominated Mobile PHR direct message (MPHR-DA).Preliminary performance tests were carried out to assess the MPHR-DA protocol with positive results.

In [31], authors focus on emergency cases. They proposed an emergency medical system through mobile cloud-based Android application with PHR. Architecture is based on Google Cloud Messaging Services.
Nowadays PHR tends to integrate information with social network, third-party applications, and other PHRs. Integration with social networks provides patients with significant benefits [32]. For instance, patients with the same disease can contact each other and share experiences [32]. In [21], researchers focus on harnessing patients social media posting to give more information to medical stakeholders.

## 3.2. PHR Recommendations For Development

Some of the filtered articles analyse and identify broad issues to certain themes in PHR system design. Notices that these articles focus giving recommendations, tips and suggestions for PHR system design.

In[2]interviews were held to three physicians to obtain characteristics and recommendations for the PHR design in the region of South Africa. Topics included questions about their hospitals' health information management. Conclusions made about these interviews addresses that physicians considered necessary inter consultation, sharing patients records among them.

Moreover, they demanded easier and faster access to medical records. This last requirement could be tackled by ubiquitous access via portable devices (e.g. smartphones, tablets).

According to the mobile survey[5], an effective PHR should provide decision support, such as medication interaction prevention feedback, for its users. The pervasiveness of mobile PHRs (mPHRs) can power decision support.

Pervasiveness of mobile provides the opportunity to interact with patients longer and more often than regular PHRs that require physical proximity to a computer. Thus, mPHRs may have a higher chance of success in adoption due to more the long-term interaction between the users and the mPHR[5].

Cloud computing represents an ideal opportunity to develop applications that ensure high performance data processing and easy management of the different tools in medical environment ensuring a consistency storage capabilities [33]. We identified two surveys about cloud computing in healthcare [33,34]. The first one details a systematic literature review in cloud computing and classified the area studies (cloud-based eHealth framework design, applications of cloud computing, and security or privacy control mechanisms of healthcare data in the cloud). The second survey article corresponds to a survey concerning the current models of health that are switching to solutions based on cloud computing [33].

Both surveys concludes that hybrid cloud is a reliable approach. Hybrid cloud services are deployed over an infrastructure that use the private mode for certain aspects (e.g., storage of data) and the public mode for other (e.g., access interfaces). Hybrid cloud ensures an efficient and robust solution for each medical and clinical department thanks to the use of between public and private aspect and advantages of the cloud distribution model [33].

As an architectural recommendation, the survey made in [1]introduces and describes two infrastructural PHR drivers: ubiquitous technology baseline for PHRs, and connectivity coverage. *Ubiquitous technology* deals with issues such as storage type (fixed or portable), software/hardware requirements and web-based infrastructure, and *connectivity coverage* deals with the physical location of the PHR data. These two can affect the PHR architecture design characteristics of availability and reliability. To provide a global explanation of both infrastructure drivers, they made a classification of PHRs depending on the storage (local, remote server-based, hybrid PHR). Later, in same work, eleven functional PHR capabilities are described, providing a basis for the analysis of the relationships between the two infrastructural drivers and architectural selection. Functional capabilities in PHRs could affect PHR adoption and usage. Moreover, it can enhance healthcare service such as improving the quality of care and safety, decision-making empowerment, patient centered and continuous care, and reduction of healthcare cost.

The key issue and requirements are given for each of eleven functional capabilities in areas such as security, data management, interoperability[35], and others [1].

As an area related to the development for ubiquitous devices, various works give recommendations in the usability and adoption arena. Some consequences of low adoption of PHR systems is due to the lack of evolving during design process to final users (patients and physicians) [36]. Moreover, according to a study made in South Africa [36], patients are interested in accessing their information to improve their healthcare. While patients are concerned about the safety of their online health information, this does not constitute a barrier to the PHR usage.

As a statistic fact, PHR is used more by people with disabilities and chronic conditions and people caring for elderly parents [37,32].Because these kinds of patients access more frequently, ubiquitous devices (e.g, smartphones or tablets) could help them to alleviate this task accessing to information day-by-day.

### 3.3. System Integration and Standards

In this research topic, we included articles that describe integration architecture in PHR systems. Moreover, standards applied to achieve interoperability are explained.

To solve the problem of data exchange between medical units, authors [7] propose interoperability through HL7 CDA (Clinical Document Architecture) Standard. CDA is a core document standard to ensure such interoperability for the purpose of exchange [38].

Interoperability between healthcare centers not only helps improve patient safety and quality care, but also reduce time and resources [39]. Another proposal applying CDA is a cloud-based CDA document generation and integration open API [39]. They integrated multiple CDA documents into a single CDA document and made it available via an open API.

An integration system is proposed in[40]. The main goal was to integrate a formal EHR care system developed at an Edinburgh Napier University (ENU, in UK) denominated DACAR with an informal care system MS HealthVault [28].

Integration is made via a *translator gateway*, and information from MS HealthVault is translated to DACAR via a secure communication channel[40]. Moreover, the *translator gateway* replicate access right from one system to another.

The system *uHospital* [29], mentioned before in Section 3.1, employs ontologies based on international and established healthcare standards as HL7 [38] and OpenEHR [41]. Applying ontologies allow to detect possible patient health risk in stored information.

In the mobile area, implementing proper standards such as messaging structures (e.g., HL7) or medical vocabularies (e.g., SNOMED, ICD10), can integrate information among EHRs, insurers, pharmacies, or consumer medical devices (e.g., glucometers, blood pressure monitor) [5].

The last but not least article in this section corresponds to a PhD thesis [42] that investigated technical challenges and barriers for integrating EMR (Electronic Medical Record) and PHR systems. First of all, the author analysed the PHR infrastructure in Netherlands focusing on interoperability standards (HL7 FHIR). Sixty-eight cloud-based PHR systems were examined in order to understand their characteristics and features. As a conclusion, only fifteen systems describe to be equipped with an API, while eight of them are public. Only two systems, those of Microsoft and Google, are positioned as health information platforms, which provide tools to developers to create applications that interact with the platform. Finally, the author proposes a prototype scenario of interoperability between an EMR system and PHR of MS HealthVault [28]}. The research concludes with an architecture that implements a core integration design called AORTA [42]. For integration, AORTA relies on HL7 FHIR standards because is an open, relatively easy and formal standard. Moreover, HL7 FHIR includes an optional RESTful API that was included in the design [42].

## 3.4. Security and Privacy

In this section we included articles that address security and privacy mechanism in PHR system. According to [43], there are three important security and privacy challenges: *privacy of the patient's data*, *access control management* and, *authenticity and integrity of data*. We will explain the three challenges and describe current works that pursue to tackle them. To ease reader lecture, this topic is divided into subsections.

### 3.4.1. Privacy of the Patient's Data

It refers to the willing of patient to divulge some of his sensitive health information. A patient can allow or deny sharing their information with a given healthcare practitioners [43].

In most cases, users who create EHRs are also responsible for generating access policies based on attributes of authorized healthcare providers [44]. In [45], data owner is responsible for generating access policies associated with its records, anonymizing and encrypting records with the corresponding policies and uploading the encrypted records into the cloud. Providing a privacy policy for PHRs should be mandatory, and constantly renewing the content and structure of the privacy policies is recommended [46].

Besides access to patient's data for clinical purpose, it is also possible to use the data for researches, which is known as secondary use of information. Users under the research role should not be provided specific details of any individual patient. This control can be achieved by anonymizing the file based on the role during decryption phase, only required data is selectively disclosed, depending on the role of the requester [45].

Another issue in research is the keyword set used during searches, this should remain hidden from the cloud server, to avoid the association of a particular research over a group of patients. In [47], the authors propose a Secure Channel Free Public-Key Encryption with Keyword Search (SCF-PEKS), which enables users to perform private searches for matching keywords over encrypted data without revealing the keywords or partial matches to the server.

To facilitate analysis and research, [48] proposes to cluster PHR using a proposed *Efficient K-Means Clustering* (EKMC) algorithm. To preserve patients' privacy, clustered PHR is anonymized using suppression and generalization techniques. To reduce the cost of cloud storage, another proposed algorithm, the Data Aggregation and Data Deduplication (DAD), is applied to remove the repeated data. Finally, PHR is encrypted through AES (Advanced Encryption Standard) encryption before outsourcing to prevent the misuse of data by unauthorized users. Another approach for health records anonymization is presented in [49], where authors propose a scalable two-phase top-down specialization (TDS) approach to anonymize large-scale data sets using the MapReduce framework on the cloud.

It is assumed that the cloud service providers are semi-trustworthy (honest but curious) [50,51,52]. This implies that cloud service providers honestly performs legitimate protocols but passively observes traffic in the cloud [50].

Encryption techniques are applied to avoid unauthorized access to the PHR data [50,51,52]. In [53] authors propose that personal health record transactions be managed using geometric data perturbation in cloud computing. Data perturbation is one of the major techniques for preserving privacy. Geometric data perturbation significantly reduces complexity in balancing data utility and guaranteeing data privacy. Geometric data perturbation (GDP) consists of a sequence of random geometric transformation. During the outsourcing, the cloud receives no details of the original private PHR data. Another approach, which is more common in the literature, is to encrypt data based on the role access policies and then outsource the data to the cloud [44]. We found that Advanced Encryption Standard (AES) was employed in several works for encryption [47,48,54,55]. There are various encryption mechanisms based on role access, this will be discussed in the next sub-Section.

### 3.4.2. Access Control Management

Access control policies should be defined and applied to the health record ensuring that information is accessible only to authorized parties [43]. One of the main challenges in this matter is to define authorization rules for each user at very fine granularity while maintaining the desired access efficiency and availability of the EHR system [43]. In [32] the authors identified eight roles and seven access levels which the PHR designers should consider, including owner, friends, family, healthcare professionals, other users, devices, applications, and other services (like insurance companies and pharmacies).

A strong access control mechanism is necessary to ensure data privacy. Access control mechanisms can be broadly classified into three categories: Identity Based Access Control (IBAC), Role Based Access Control (RBAC), and Attribute Based Access Control (ABAC). IBAC is not feasible in clouds because the number of users in the cloud environment is too large to keep them in the Access Control List (ACL) that contains the identity of authorized users [56]. According to [57], after comparison made between RBAC and ABAC, the latter model has stronger ability to express complex access policies than RBAC model. Moreover, RBAC-based rule can be converted and integrated into ABAC-based policy. In ABAC, users are tagged with certain attributes, and the data has attached access policies. Only the user with a valid set of attributes that satisfy the access policies, can access the data. Most of the existing works on ABAC-based access control mechanisms use a cryptographic primitive Attribute Based Encryption (ABE) [56]. In [43], the authors recommended to adopt a combination of Role-Based Access Control (RBAC) [43] and Attribute-Based Access Control (ABAC) approaches as access control management.

In ABE one can embed an access policy into the ciphertext or decryption key in ABE system. Compared with traditional public key encryption and identity-based encryption, ABE has a significant advantage as it achieves flexible one-to-many encryption instead of one-to-one. Moreover, data access is self-enforcing from the cryptography, requiring no trusted mediator [58]. Many works that propose similar approach, by encrypting PHR files before upload them to the cloud, use ABE approaches, among them: [53,59,60,61,62,63,64,65,66,67], [68,69,70,71,72]. In [58] the authors introduce a Python ABE library named *pyabelib*.

There are several variations of ABE, two of them are of the most spread use: Key-policy Attribute-based Encryption (KP-ABE) and Ciphertext-policy Key-policy Attribute-based Encryption (CP-ABE). Several proposal uses only CP-ABE such as[44,45,47,50] as encryption mechanism.

In [51] key-policy attribute-based encryption (KP-ABE) and ciphertext-policy attribute-based encryption (CP-ABE) are used together, this schemes encryption are conceptually similar to ABAC and RBAC respectively. The essential difference between this two schemes is whether the system use attributes to describe the encrypted data or the user's private key. Generally speaking, CP-ABE is conceptually closer to RBAC, while KP-ABE is closer to ABAC [51].

The main advantages of combining KP-ABE and CP-ABE is the possibility to conform two different attributes sets. In [51], the authors define a date-based attributes set using medical-specific attributes, such as surgical profile, medication profile, cardiac profile, etc. for KP-ABE, while for CP-ABE, define a role-based attribute set using the role information such as *john.doe.family*, *john.doe.friend* and *john.doe.physician*. As result, the attributes are grouped into two domains, the public domain which refers to the intrinsic properties of PHR data, and the personal domain which refers to the personal identifiable information of the entities in the PHR system [51]. For KP-ABE, each access structure *A* defines the privilege of individual medical personnel. For CP-ABE, each access structure *A* determines who can access a particular PHR file [51].

Other works such as [52,73]} also use the two domain approach (public and personal) to facilitate key distribution. Users within Public Domain (PUD) can have access through their professional roles, and users within the PerSonal Domain (PSD) can be reduced to people associated with the patient (such as family, personal caregiver, closest friends). Patients can specify fine role-based fine-grained access policies for his PHR files. With this, patients do not need to know a list of authorized users once the policy is defined.

Another complexity regarding access control covers emergency scenarios. Timely availability of medical data needs to be guaranteed, especially under emergency cases[32,43], thus, special access control policies can be applied under this particular circumstance. Considering emergencies scenarios, where an Emergency Department can provide temporary read-access to medical staff, and when the emergency is over, PHR owner may revoke this emergency access [52].

Among others ABE variations are: Hierarchical Attribute-Set-Based Encryption (HASBE) [74], online/offline ciphertext-policy attribute-based proxy re-encryption scheme (OO-CP-AB-PRE) [75], named attributes ABE [76], multi-authority ciphertext-policy ABE [77], hierarchical identity-based encryption (HIBE) [55], Multi Authority ABE [78].

The Hierarchical Attribute-Set-Based Encryption (HASBE) extends the cipher text-policy Attribute-Set-Based Encryption (ASBE) with a hierarchical structure of users utilizing compound attributes [74]. The intended scheme achieves fine-grained, flexible and scalable data access control with the help of compound attributes of HASBE. In [79,80] the HASBE approach is also applied to a PHR system, in these cases an attribute concerning expiration time is added to the keys.

Ciphertext-policy attribute-based proxy re-encryption scheme (OO-CP-AB-PRE) scheme supports Attribute-Based Proxy Re-Encryption with any monotonic access structures and deals with Online/Offline encryption [75].

In named attributes ABE, each attribute contains an attribute name and its value. For the decryption process, even though attribute names of the receiver meet the access, it should first check all possible attribute values to have access to the data. To overcome computation overhead, authors adopt the Bloom filter [76].

In multi-authority ciphertext-policy ABE scheme, the access policy is hidden and hence user access privacy is protected. A PHR user obtains his attribute private key and if the attribute set associated with the private key does not satisfy the access policy in a PHR ciphertex, PHR user cannot decrypt and guess what access policy was specified by the PHR owner. Hence, access policy is hidden, and user access privacy is protected [77].

The multi-authority attribute based encryption scheme (MA-ABE) includes multiple attribute authority for handling the different set of users from various domains [78]. This approach can be used together with others ABE techniques where a Trusted Authority is needed.

A survey on different encryption schemes using ABE was done in [78], it shows that the Multi-Authority Attribute Based Encryption (MA-ABE) got the highest score in the comparison between several ABE based techniques.

There are others approaches for role access management. In [81]the authors propose an authorization policy built in Ontology Web Language (OWL) and ontology rules defined in Semantic Web Rule Language (SWRL).

In [82], a bilinear pairing approach for access control to PHR's data is presented. A certificate authority (CA) stores a matrix access [U, D], where each Ui represents a User and each Di represents the data. If the cell has the value 1, then the User Ui is allowed to access Data Di. The CA encrypt each Di and provides the Decryption Key (DKi) to the users who are allowed to access the data.
An access control scheme based on Lagrange interpolation polynomial under Cloud computing environments is proposed in [83]. The proposed scheme can resist internal and external attacks, is convenient for a Central Authority to manage by using only one public formula G(x, y), the generation of keys and the algorithms are simple, the PHR system allows patients to determine the access users and remove the outdated authorization.

For all the approaches discussed above, the data is stored encrypted, thus efficiency in keyword search is important. For this purpose, authors in [71] propose a privacy-aware fuzzy keyword approach. Another important issue is to ensure the integrity and authenticity of data of PHR records, the next sub-Section describes works related in this field.

### 3.4.3. Authenticity And Integrity Of Data

It concerns to the fact that data has not been manipulated by unauthorized use. Data must remain consistent [43]. Digital signature is a very useful tool for providing authenticity and integrity [43]. Trust authorities are also very useful, a trust authority is usually an organization expected to be responsible for supervising healthcare information exchange among healthcare system. System build upon cryptography scheme, usually employ a trust authority as the responsible for distributing decryption keys to corresponding medical personnel [44,45,51,84].

In [85], an encryption scheme called DMACPSABE provides extensions to the Ciphertext-Policy Attribute-Based Encryption model to add multiple distributed authorities which share a subset of attributes without the need for a user to communicate with more than one authority. In the two domains approach, PUblic Domains (PUDs) and PerSonal Domains (PSDs), each PHR owner is a trusted authority of his own PHR, so the patient himself can give access to particular users in the PSD [52]. In [47], with the patient-controlled encryption approach, each owner will be able to generate keys for the ABE scheme, without relying on an outside trusted authority.

To integrate data from different providers, the EHR records should first be verified. In [43], the *EHR secure collection and integration* component has the task of verifying EHRs provided by different healthcare centers, in terms of authenticity, confidentiality, and integrity to combine and integrate the successfully verified EHR data into a new composite EHR. This work recommends the use of anonymous digital credentials.

The main idea behind anonymous digital credentials is that users are given cryptographic tokens which allow them to prove statements about themselves and their relationships with public and private organizations anonymously. Such credentials, while still making an assertion about some property, status, or right of their owner, do not reveal the owner's identity [43]. According to [12,43], audit and archiving are two alternative security metrics to measure and ensure the safety of a healthcare system. Audit can be performed by maintaining a log of every access to and modification of data. Archiving means moving healthcare information to off-line storage in a way that ensures the possibility of restoring them to on-line storage whenever it is needed without the loss of information [43].

Others proposals to ensure the authenticity and integrity of data are: Identity Provider [84], Selective Anonymity for a Privacy Preserving Health Information REpository (SAPPHIRE) [86], Ciphertext-Policy Attribute-Based Signcryption (CP-ABSC) [87]and data access through Trusted Authorities [55].

This section focus in the variants of the ABE security mechanism. The ABE approach covers the patient's privacy and access management areas, by the leverage of granularity permission to disclose data and respecting patient's privacy policies. The division of users that can get access to the patient data in the PerSonal and PUblic Domains (PUD and PSD) simplifies the management of the access permissions. Finally, the distribution of the attributes keys among several trusted providers, each one in charge of a specific set of attributes according a context, improves reliability in the distribution of tokens by avoiding bottlenecks and enhancing security in the process.

## 4. ARTICLES STATISTICS AND DISCUSSIONS

As a statistic resume, 101 articles were included in the final synthesis. In Figure 2 quantity of articles found each research topic is showed in the end of columns. Some articles may be explained in more than one topic, for this analysis we counted articles just once. Therefore, an article was included in the research topic that describes its core information the most. Figures percentages are summarized as follows:

A. Proposal/Developed systems: 35 articles (34,7%).
B. PHR recommendations for development: 11 articles (10,8%).
C. Systems integration and standards: 4 articles (4%).
D. Security and Privacy: 51 articles (50,5%).

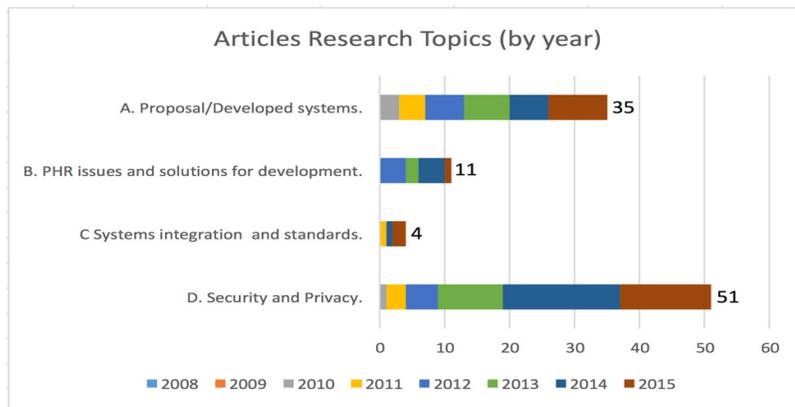

Figure 2. Articles research topics (classified by year) with quantities

More than 50% of the articles identified corresponds to *security and privacy* (Topic D) mechanism. Even though our primary focus was in PHR system that includes security and privacy mechanism (Topic A) publications, we identified more articles in *security and privacy* concerns.

As the review question stays, our main focus in this work is to identify related works PHR system architectures that apply three characteristics in their design: integrated, reliable and cloud-based. The main research topic *proposal/developed systems* (Topic A) addresses our research goals. Thirty-five articles explain proposal or developed PHR systems that applied the characteristics. The most common approach applied by architectures found is SOA (Service-oriented architecture).

Even though the primary focus of this study was already in *proposal/developed systems* (Topic A), our research stage filtered other articles that deal with the one or more characteristics partially.

The term *reliable* corresponds to articles identified in *security and privacy*(Topic D). As explained above most of the articles filtered in the scoping review were from security and privacy.

The term *integrated* is boarded in research topic *System integration and standards* (Topic C). Most articles applied HL7 CDA (Clinical Document Architecture) standard to ensure interoperability. It consists of a markup standard to specify structure semantic of clinical document [39]. The first version was developed in 2001, and the second release, in 2005. Although there is nothing wrong applying older standard, newer authors in the community have started to test a newer standard as HL7 FHIR (2014) in their architectures as in [42].

Regarding the last term *cloud-based*, this review focused on searching cloud architecture PHR beforehand. Besides taking advantage from other benefits from cloud technology (scalability, elasticity, pay-on-demand, among others), our main purpose was to achieve *ubiquity* to ease information access. As a recommendation, two surveys holds that hybrid cloud is an ideal approach for healthcare system because of their efficiency and robustness when mixing public and private cloud services [33,34]. Third-party cloud services (e.g., Amazon S3, Google Cloud, Microsoft Azure) are also used by PHR. In this context, *cloud* technology is addressed in this topic research (Topic A).

Table 1. Characteristics and research topics relationship.

| Characteristics | Research topics |
|---|---|
| Reliable | PHR recommendations for development (Topic B), and Security and privacy (Topic D). |
| Integrated | PHR recommendations for development (Topic B), and System integration and standards (Topic C). |
| Cloud-based | Proposal/Developed systems (Topic A), and PHR recommendations for development (Topic B). |

Research topic *PHR recommendations for development*(Topic B) gives recommendations for a variety of areas (e.g., usability and adoption, PHR information recommendations, cloud computing). For this reason, we defined that Topic B is related to the three terms. Characteristics searched and their relationship to researched topics found are summarized in Table 1.

We focused our research in the privacy aspects of PHR systems, the resulting set of works found in this area also involve other informatics security mechanism, such as role management, encryption, authentication, etc. As can be seen in Figure 2, from articles identified in *security and privacy*(Topic D), most of them correspond to years: 2013 (19,6%), 2014 (35,3%), and 2015 (27,5%). This figure gives as a result that more than 80% of the articles in this research topic were published since 2013. This shows how novel this subject is in the field, being ABE mechanism the greatest focused approach.

The second topic with most articles belongs to *Proposal/Developed systems*(Topic A). Articles are distributed over years of publications as follows: 2010 (8,6%), 2011 (11,5%), 2012 (17,1%), 2013 (20,0%), 2014 (17,1%), and 2015 (25,7%). We can infer that PHR development has been increasing during years, and being also quite research area. Since 2013, 65,6% of articles in this Topic A were published.

The other two research topics left, together sum 14,9%; for *PHR recommendations for development* (Topic B) 10,9%, and for *Systems integration and standards*(Topic C) 4% (See Figure 2).

In Table 2 we present a reference summary of articles included and the research topics where they where included. Notice that some articles are repeated into three topics at most.

Table 2. Articles references with research topics.

| Ref | A | B | C | D | Ref | A | B | C | D | Ref | A | B | C | D | Ref | A | B | C | D | Ref | A | B | C | D |
|---|---|---|---|---|---|---|---|---|---|---|---|---|---|---|---|---|---|---|---|---|---|---|---|---|
| [1] | | | X | | [23] | X | | | | [47] | | | | X | [67] | | | | X | [87] | | | | X |
| [2] | | | X | | [24] | X | | | | [48] | | | | X | [68] | | | | X | [88] | | | | X |
| [3] | X | | | | [25] | X | | | | [49] | | | | X | [69] | | | | X | [89] | | | | X |
| [4] | X | | | | [26] | X | | | | [50] | | | | X | [70] | | | | X | [90] | | | | X |
| [5] | | X | | | [27] | X | | | | [51] | | | | X | [71] | | | | X | [91] | | | | X |
| [6] | X | | | | [29] | X | | X | | [52] | | | | X | [72] | | | | X | [92] | | | | X |
| [7] | X | | X | | [30] | X | | | | [53] | | | | X | [73] | | | | X | [93] | | | | X |
| [9] | X | | | | [31] | X | | | | [54] | | | | X | [74] | | | | X | [94] | | | | X |
| [10] | | X | | | [32] | X | X | | X | [55] | | | | X | [75] | | | | X | [95] | | | | X |
| [11] | X | | | | [33] | | X | | | [56] | | | | X | [76] | | | | X | [96] | | | | X |
| [12] | | | | X | [34] | | X | | | [57] | | | | X | [77] | | | | X | [97] | | | | X |
| [13] | X | | | | [36] | | X | | | [58] | | | | X | [78] | | | | X | [98] | | | | X |
| [14] | X | | | | [37] | | X | | | [59] | | | | X | [79] | | | | X | [99] | | | | X |
| [15] | X | | | | [39] | | | X | | [60] | | | | X | [80] | | | | X | [100] | | | | X |
| [16] | X | | | | [40] | | | X | | [61] | | | | X | [81] | | | | X | [101] | | | | X |
| [17] | X | | | | [42] | | | X | | [62] | | | | X | [82] | | | | X | [102] | | | | X |
| [18] | X | | | | [43] | | | | X | [63] | | | | X | [83] | | | | X | [103] | | | | X |
| [19] | X | | | | [44] | | | | X | [64] | | | | X | [84] | | | | X | [104] | | | | X |
| [20] | X | | | | [45] | | | | X | [65] | | | | X | [85] | | | | X | [105] | | | | X |
| [21] | X | | | | [46] | | | | X | [66] | | | | X | [86] | | | | X | [106] | | | | X |
| [22] | X | | | | | | | | | | | | | | | | | | | | | | | |

## 5. CONCLUSION

The goal in the scoping review was to identify current implemented or proposed PHR systems that achieve three characteristics: integrated, reliable and cloud-based. In general, we identified thirty-five articles that achieve the three characteristics. However, more articles that partially achieve the cited characteristics were found. After searching and filtering stages, we included 101 articles in our final synthesis. We identified four research topics areas: proposal/developed systems, PHR recommendations for development, systems integration and standards, and security and privacy.

Regarding each characteristic stayed in our review question, ABE mechanism is the most used approach for fulfilling *reliability*. For *cloud-based*, the majority of systems proposed access to cloud services applying SOA via web servers which ease access for mobile devices. Then, hybrid cloud is addressed as the ideal approach for PHR deployment. For last researched term *integration*, most works applied HL7 CDA (Clinical Document Architecture) as their standard solution. Moreover, HL7 FHIR is a standard that is blooming out in newer proposals.
A challenging future work would be to propose a PHR system architecture that accomplishes the three key features (integration, reliable and cloud-based) with novel standards in compliance with nowadays technology.

## ACKNOWLEDGEMENTS

This work was funded by "Programa Paraguayo de Ciencia y Tecnología" of CONACyT (*Consejo Nacional de Ciencia y Tecnología*). Project adjudicated in 2013 by Prociencia call 14-INV-471.

Health Informatics - An International Journal (HIIJ) Vol.5, No.2/3, August 2016